\documentclass[twocolumn]{aastex6}
\usepackage{natbib}
\usepackage{amsmath}
\usepackage{graphicx}
\usepackage{hyperref}

\shorttitle{Dark Matter in Clusters from Radio Emission}

\shortauthors{Storm, Jeltema, Profumo}

\begin{document}

\title{Synchrotron Emission from Dark Matter Annihilation: Predictions for Constraints from Non-detections of Galaxy Clusters with New Radio Surveys}

\author{Emma Storm\altaffilmark{1,2}, Tesla E. Jeltema\altaffilmark{3,4}, Megan Splettstoesser\altaffilmark{3}, Stefano Profumo\altaffilmark{3,4}}
\altaffiltext{1}{GRAPPA, Universiteit van Amsterdam, Science Park 904, 1098XH Amsterdam, The Netherlands}
\altaffiltext{2}{e.m.storm@uva.nl}
\altaffiltext{3}{Department of Physics,  University of California, 1156 High St., Santa Cruz, CA 95064, USA}
\altaffiltext{4}{Santa Cruz Institute for Particle Physics,  University of California, 1156 High St., Santa Cruz, CA 95064, USA}


\begin{abstract}

The annihilation of dark matter particles is expected to yield a broad radiation spectrum via the production of Standard Model particles in astrophysical environments. In particular, electrons and positrons from dark matter annihilation produce synchrotron radiation in the presence of magnetic fields. Galaxy clusters are the most massive collapsed structures in the universe, and are known to host $\sim\mu$G-scale magnetic fields. They are therefore ideal targets to search for, or to constrain the synchrotron signal from dark matter annihilation. In this work we use the expected sensitivities of several planned surveys from the next generation of radio telescopes to predict the constraints on dark matter annihilation models which will be achieved in the case of non-detections of diffuse radio emission from galaxy clusters. Specifically, we consider the Tier 1 survey planned for the Low Frequency Array (LOFAR) at 120 MHz, the EMU survey planned for the Australian Square Kilometre Array Pathfinder (ASKAP) at 1.4 GHz, and planned surveys for APERTIF at 1.4 GHz. We find that, for massive clusters and dark matter masses $\lesssim 100$ GeV, the predicted limits on the annihilation cross section would rule out vanilla thermal relic models for even the shallow LOFAR Tier 1, ASKAP, and APERTIF surveys. 

\end{abstract}


\section{Introduction}

While astronomical observations have revealed much about the amount and macroscopic properties of dark matter in our universe, we still do not know what particle or particles constitute the dark matter, nor what beyond the Standard Model particle physics it points to.  A leading class of dark matter particle candidates are weakly interacting massive particles (WIMPs) \citep{1996PhR...267..195J, 2000RPPh...63..793B, 2005PhR...405..279B}.  WIMPs may be thermally produced in the early universe with the right relic density to explain the present dark matter density if they have an annihilation cross section of $\langle \sigma v \rangle \sim 2.2 \times 10^{-26}$ \citep[e.g.][]{2012PhRvD..86b3506S}.  The pair annihilation of WIMPs to Standard Model particles in dark matter dense structures would then give a broad range of potentially observable signatures; the expected products of WIMP annihilation include gamma-ray photons, high-energy electrons and positrons, and neutrinos.  Many recent indirect dark matter searches have focused on potential gamma-ray emission from dark matter annihilation (see, e.g., \citealt{Charles2016,Conrad2015,Porter2011,Feng2010} for reviews), and in particular, some of the strongest limits on dark matter annihilation for supersymmetric dark matter models come from the non-detection of local dwarf spheroidal galaxies with the Fermi Large Area Telescope (Fermi-LAT) \citep{Ackermann2015}. However, dark matter annihilation generically  produces emission across the entire electromagnetic spectrum \citep[e.g.,][]{Colafrancesco2006, Profumo:2010ya, Profumo:2013yn}.  For example, in the presence of magnetic fields, the high-energy electrons/positrons produced will radiate via synchrotron in radio.  These same high-energy particles  inverse-Compton (IC) up-scatter background radiation, such as the cosmic microwave background (CMB) or starlight, to X-ray or soft gamma-ray frequencies.

Clusters of galaxies represent particularly promising targets for searches for secondary IC and synchrotron emission resulting from dark matter annihilation or decay \citep[e.g.,][]{Colafrancesco2006, 2012MNRAS.421.1215J, Storm2013}.  In addition to being dark matter dominated, clusters contain large-scale magnetic fields, and the sheer size of clusters enables them to confine $e^\pm$ produced by dark matter long enough for them to radiate.  In \cite{Storm2013}, we demonstrated the potential of radio observations of clusters for dark matter studies: the non-detection of radio emission from nearby galaxy clusters strongly constrains the dark matter annihilation cross-section, and in many cases radio observations place better limits than current gamma-ray cluster data.

Some clusters are observed to host Mpc scale diffuse radio synchrotron emission in the form of radio halos or radio relics, but most do not \citep[e.g.][]{2012A&ARv..20...54F, 2014IJMPD..2330007B}. This radio emission is thought to stem from high-energy electrons accelerated in cluster mergers.  For example, radio halos are primarily observed in disturbed clusters and may result from the turbulent acceleration of mildly relativistic electrons in cluster mergers \citep[e.g.][]{2010ApJ...721L..82C, 2014IJMPD..2330007B, 2016MNRAS.458.2584B}.  A second hypothesis is that radio halos are hadronically generated in the collisions of cosmic ray protons with the intracluster medium. Such possibility is however disfavored by the non-detection of gamma-ray emission from clusters and by the spatial distribution of the radio emission \citep[e.g.][]{2011ApJ...728...53J,2012MNRAS.426..956B, Zandanel2013, 2015MNRAS.448.2495S}.  In setting limits on dark matter models, one is free to choose target clusters which lack contaminating radio signals from either central AGN or radio halo/relic emission.  A radio signal from dark matter, on the other hand, would appear as diffuse radio emission with a consistent spectrum across targets, and with a brightness scaling with cluster mass rather than cluster dynamical state or other factors.

Radio astronomy is entering a new era with several new observatories coming online recently or in the near future, and offering dramatically increased sensitivity, new low frequency capabilities, and improved spatial resolution.  The spectrum expected for dark matter annihilation/decay rises sharply at low frequencies, particularly for low dark matter particle masses, making low-frequency radio observations ideal for dark matter searches.  At the same time, the Square Kilometer Array pathfinders will provide novel sensitivity at GHz frequencies.  

In this paper, we investigate the sensitivity of current and near-term radio surveys to dark matter annihilation showing that these have the potential to surpass all current indirect detection limits for a large range of particle masses and annihilation final states.  In particular, we focus on planned surveys with the Low Frequency Array (LOFAR) \citep{vanHaarlem2013}, the Australian Square Kilometer Array Pathfinder (ASKAP) \citep{Johnston2008}, and a new Phased Array Feed system, APERTIF, which will be installed on the Westerbork Synthesis Radio Telescope (WSRT) \citep{Verheijen2008}.

This paper is organized as follows. In Section~\ref{sec:synch}, we derive the signal from synchrotron emission due to dark matter annihilation and describe the models for the dark matter spatial profile and magnetic fields in clusters we employ here. In Section~\ref{sec:rad}, we briefly summarize the objectives and relevant details of the upcoming surveys for ASKAP, APERTIF, and LOFAR. In Section~\ref{sec:pred}, we dicuss how we estimate upper limits for these surveys. We also present our predictions for constraining dark matter annihilation by non-detections of clusters with these upcoming surveys. Finally, we summarize our findings in Section~\ref{sec:end}. Throughout the paper, we assume a $\Lambda$CDM cosmology with H$_0=70~$km~s$^{-1}$~Mpc$^{-1}$, $\Omega_m=0.27$, and $\Omega_{\Lambda}=0.73$.

\section{Synchrotron Emission from Dark Matter Annihilation}\label{sec:synch}
\subsection{Derivation of Synchrotron Signal}

The annihilation of dark matter results in the production of a wide variety of Standard Model particles, including electrons (and positrons). We use DarkSUSY \cite{Gondolo2004} to calculate the electron injection spectrum per dark matter annihilation event. The electron injection spectrum, $\frac{dN}{dE}_{inj}$, depends on the mass of the dark matter particle and on the annihilation channel. Generally, the injection spectra cutoff sharply at the dark matter particle mass at high energies. The injection spectrum from annihilation to muons is very hard, while the spectra from annihilation to tau particles is softer, and the spectra from annihilation to quarks is softer still (see, e.g., Figure 4 in \citealt{Gaskins2016}).

Diffusion and energy losses in astrophysical environments modify the injection spectrum from dark matter annihilation. The electron equilibrium spectrum is obtained from the following diffusion equation that takes these mechanisms into account:
\begin{equation}\label{eq:diff}
  \begin{split}
    \frac{\partial}{\partial t} \frac{dn_e}{dE} &= \nabla \left[D(E,r)\nabla \frac{dn_e}{dE} \right]\\
    & + \frac{\partial}{\partial E}\left[b_{loss}(E,r,z)\frac{dn_e}{dE} \right] + Q(E,r).
  \end{split}
\end{equation}
Here, $dn_e/dE$ is the electron equilibrium spectrum, $Q(E,r)$ is the source term, $D(E,r)$ is the spatial diffusion coefficient, and $b_{loss}(E,r)$ is the energy loss term described below:
\begin{equation}
  \begin{split}
    b_{loss}(E,r,z) &= b_{syn} + b_{IC} + b_{brem} + b_{coul} \\
    &\approx0.0254\left(\frac{E}{1\mathrm{GeV}}\right)^2\left(\frac{B(r)}{1 \mu \mathrm{G}}\right)^2\\
    &+ 0.25\left(\frac{E}{1\mathrm{GeV}}\right)^2(1+z)^4\\
    &+ 1.51n(0.36+\mathrm{log}(\gamma/n))\\
    &+ 6.13n(1+\mathrm{log}(\gamma/n)/75.0).
  \end{split}
\end{equation}
The energy loss term $b_{loss}(E)$ has units of $1\times10^{-16}$~GeV s$^{-1}$, where $E=\gamma m_ec^2$ is the energy of a single electron, and $n$ is the average thermal electron density, $\approx1\times 10^{-3}$~cm$^{-3}$ for clusters. For GeV electrons and positrons, synchrotron and IC losses dominate. When $B > B_{CMB}$, synchrotron losses dominate over IC losses, where $B_{CMB}\simeq 3.25(1+z)^2 \mu$G.

The source term is related to the electron injection spectrum from dark matter self-annihilation:
\begin{equation}
Q(E,r)=\frac{\langle \sigma v \rangle \rho_{\chi}^2(r)}{2 m_{\chi}^2}\frac{dN}{dE}_{inj},
\end{equation}
where $m_{\chi}$ is the dark matter mass (in units of energy), $\langle \sigma v \rangle$ is the annihilation cross section and $\rho_{\chi}(r)$ is the spatial distribution of dark matter. 

In cluster environments, energy losses from synchrotron radiation and IC scattering dominate, and the diffusion timescale is long compared to these energy loss timescales \citep[e.g.,][]{Colafrancesco2006}. We therefore neglect the dependence on spatial diffusion and the time dependence in Equation~(\ref{eq:diff}). The expression for the electron equilibrium spectrum then reduces to:
\begin{equation}
\frac{dn_e}{dE}(E,r,z) = \frac{\langle \sigma v \rangle \rho_{\chi}(r)^2}{2 m_{\chi}^2 b_{loss}(E,r,z)}\int_E^{m_{\chi}} \mathrm{d}E' \frac{dN}{dE'}_{inj}.
\end{equation}

The synchrotron power per frequency radiated by a single electron in the presence of a magnetic field is as follows:

\begin{equation}\label{eq:Pelec}
  P_{\nu}(\nu,E,r,z) = \int_0^{\pi}\mathrm{d}\theta \frac{\mathrm{sin}\theta}{2} 2\pi \sqrt{3} r_0 \nu_0 m_e c \mathrm{sin}\theta F\left(\frac{x}{\mathrm{sin}\theta}\right) ,
\end{equation}
where $r_0=e^2/(m_ec^2)$ is the classical electron radius, $\theta$ is the pitch angle, $\nu_0 = eB/(2\pi m_e c)$ is the non-relativistic gyrofrequency (where the electron mass has units of energy). The quantities $x$ and $F$ are defined as follows:
\begin{equation}\label{eq:x}
x \equiv \frac{2\nu(1+z)}{3\nu_0 \gamma^2}
\end{equation}
\begin{equation}\label{eq:F}
    F(s) \equiv s\int_s^{\infty} \mathrm{d}\xi K_{5/3}(\xi) \approx 1.25s^{1/3}\mathrm{exp}(-s)[648+s^2]^{1/12},
\end{equation}
where $K_{5/3}(\xi)$ is the modified Bessel function of order $5/3$.

Given the equilibrium energy spectrum of electrons, the synchrotron emissivity is:
\begin{equation}\label{eq:jv}
j_{\nu}(\nu,r,z)= 2 \int_{m_e c^2}^{m_{\chi}}  \mathrm{d} E \frac{dn_{e}}{dE}(E,r,z) P_{\nu}(\nu,E,r,z).
\end{equation}

The factor of $2$ is to account for both electrons and positrons. The surface brightness is the line-of-sight (los) integral of the emissivity, and the flux density is the surface brightness integrated over the solid angle on the sky of the emission region. For small regions (where the ratio of its size to its distance is much less than one), this equation for flux density can be approximated as follows:

\begin{equation}\label{eq:ssynap}
S_{\nu}(\nu,z) \approx \frac{1}{D_A^2}\int_0^R \mathrm{d}r r^2 j_{\nu}(\nu,r,z),
\end{equation}
where $S_{\nu}$ is the integrated flux density (measured typically in Janskys), $D_A$ is the angular diameter distance, and $R$ is the maximum radius of the region of interest.

\subsection{Dark Matter Density Profile}

We assume a Navarro-Frenk-White (NFW) model for the dark matter halo density profile \citep{Navarro1996, Navarro1997}:
\begin{equation}\label{eq:rhoNFW}
  \rho_{NFW}(r) = \frac{\rho_s}{\frac{r}{r_s}\left(1+\frac{r}{r_s}\right)^2},
\end{equation}
where $\rho_s$ is the central density and $r_s$ is the characteristic scale radius. The central density $\rho_s$ is determined from scale radius $r_s$, and the virial mass and radius of the object of interest \citep{Hu2003}:
\begin{equation}\label{eq:rhos}
  \rho_s = \frac{M_{vir}}{4 \pi \boldsymbol{r_{vir}^3} f(r_s/r_{vir})},
\end{equation}
where 
\begin{equation}\label{eq:NFWint}
  f(x) = x^3[\mathrm{ln}(1+x^{-1})-(1+x)^{-1}].
\end{equation}

We use the scaling relationship derived from X-ray observations of clusters \citep{Buote2007} to determine $r_s$ from the virial mass and radius:
\begin{equation}\label{eq:cM}
  c_{vir} = 9\left(\frac{M_{vir}}{10^{14}h^{-1}}\right)^{-0.172},
\end{equation}
where the concentration $c_{vir} = r_{vir}/r_s$. We derive the virial mass and radius of the Coma cluster from the $M_{500}$ and $r_{500}$, as determined by X-ray observations \citep{Chen2007}, and corrected for our adopted cosmology. The concentration-mass relation is in general redshift-dependent. However, we choose to keep the density profile and total mass fixed, and simply calculate the resulting synchrotron signal from this dark matter halo profile at different redshifts. This is because we want to focus on just how the signal varies as redshift, without introducing any evolution of the dark matter halo that would substantially complicate the redshift-dependence. Of course, for actual observations of clusters, the full, redshift-dependent calculation for the dark matter density parameters should be used.

The NFW profile is a conservative choice in that it is a smooth profile with no substructure. There certainly exist dark matter subhalos down to the mass scale of at least dwarf galaxies ($10^7 M_{\odot}$) in clusters. However there are only small boosts ($\sim10\%$) to the synchrotron signal when a profile with this level of substructure is used \citep{Storm2013}. Integrating down to subhalo masses of $10^{-6}M_{\odot}$ gave boost factors of as much as a factor of $2$ in \citep{Storm2013}, but the precise value depends on the profile shape of the substructure and the region of interest considered.

We only consider here smooth, azimuthally symmetric dark matter distributions. Mass distributions in clusters can be considerably more complicated than this, especially in merging clusters with more than one subcluster peak; the prototypical example of such a cluster is the Bullet Cluster \citep{Clowe2006}. The most accurate estimate of the synchrotron signal from such clusters would require density profiles specific to those objects. A single NFW profile is most appropriate for clusters that are more relaxed. Coma is in fact a merging cluster; however, a NFW profile remains a good fit to its dark matter density inferred from weak lensing \citep[e.g.][]{Lokas2003}.

\subsection{Magnetic Field Modeling}\label{sec:magfield}

The detection of synchrotron radio halos in clusters proves the existence of magnetic fields distributed throughout the cluster volume. Inferred magnetic field strengths from Faraday Rotation measures are typically in the few to tens of $\mu$G range \citep[e.g.,][]{Carilli2002}. An extensive study of the magnetic field in the Coma Cluster by \citet{Bonafede2010} found that the radial profile of the field was best fit by one that scales with a beta model fit to the gas density in the cluster. Faraday rotation studies and simulations of other clusters also support this scaling \citep{Murgia2004,Vacca2012}.

We therefore assume a magnetic field profile that scales with a beta-model fit to the thermal gas density in a cluster:
\begin{equation}\label{eq:Br}
  B(r) = B_0\left(1+\left(\frac{r}{r_c}\right)^2\right)^{-3/2\beta \eta}.
\end{equation}

From Faraday Rotation measures, cool-core, relaxed clusters appear to host magnetic fields that are more centrally peaked than merging, unrelaxed clusters \citep{Carilli2002,Taylor2006}. Given this data, we  consider two sets of parameters for the magnetic field: a ``cool-core''-strength field and a ``non-cool-core''-strength field. We set $B_0=25$~$\mu$G and $r_c=46$~kpc for our ``cool-core'' model. This choice of $B_0$ is the inferred central magnetic field value from Faraday rotation measures of the prototypical cool-core cluster Perseus \citep{Taylor2006}. The value for $r_c$ is the best-fit core radius for a beta-model fit to the thermal gas density of Perseus derived from X-ray observations and corrected for our adopted cosmology \citep{Chen2007}. For our ``non-cool-core'' model, we set $B_0=4.7$~$\mu$G and $r_c=291$ kpc. This value for $B_0$ is inferred from Faraday rotation measures of the Coma Cluster \citep{Bonafede2010}. The value for $r_c$ is similarly the best-fit value for a beta-model fit to the thermal gas density of the Coma Cluster used in \citet{Bonafede2010}, derived from X-ray observations in \citet{Briel1992}. For both models, we choose $\eta=0.5$ and $\beta=0.75$, following \citet{Bonafede2010}. Varying $\beta$ and $\eta$ from $0.5-1$ changes our results by at most $40\%$ at low dark matter masses, annihilation to quarks, and for the ``cool-core'' magnetic field profile. For the ``non-cool-core'' model, the effect of $\beta$ and $\eta$ is smaller, down to $<1\%$ for annihilation to muons and $m_{\chi}=1000$~GeV.

\section{Upcoming Radio Surveys}\label{sec:rad}

There are several new and planned radio telescopes being built around the world, partially to test new technologies and observational strategies for the upcoming SKA. We choose to focus here on three instruments that span several decades in frequency space and are either actively being built or are already operational, and on four planned surveys by these instruments that have clearly delineated, and complementary, observational strategies.

\subsection{The EMU Survey with ASKAP}

The Australian Square Kilometre Array Pathfinder (ASKAP), is an interferometer located in the future headquarters of the Australian part of the SKA. It consists of 30 12-meter telescopes in a configuration that balances the need for high resolution and enhanced sensitivity to diffuse emission. The primary science goal of ASKAP is to produce deep surveys of the southern sky in the $\sim$GHz frequency range \citep{Johnston2008}. One such survey, the Evolutionary Map of the Universe (EMU), is planned to survey $75\%$ of the sky, up to a declination of $+30^{\circ}$, at $1.4$~GHz to a depth of $10~\mu$Jy and resolution of $10$~arcsec \citep{2011PASA...28..215N}. EMU is expected to begin survey operations in 2017\footnote{\url{http://www.atnf.csiro.au/people/Ray.Norris/emu/}}. 

\subsection{Surveys with APERTIF}

The Aperture Tile in Focus (APERTIF) system is an upgrade to the existing Westerbork Synthesis Radio Telescope (WSRT) in the Netherlands that will provide WSRT with an instantaneous field of view of about $8^{\circ}$, roughly a factor of 30 larger than the current FOV. This will allow APERTIF to perform fast, deep surveys \citep{Verheijen2008}. 
Two of the planned surveys include a Shallow Northern Sky Survey (SNS) with a depth of $\sim 13~\mu$Jy/beam and a Medium Deep Survey (MDS) covering $\sim 350$ deg$^2$ to $\sim 8~\mu$Jy/beam with a resolution of 15-35~arcsec\footnote{\url{https://www.astron.nl/sites/astron.nl/files/cms/Apertif_Draft_Surveys_v2.4.pdf}}.  These surveys are expected to start in 2017 and are highly complementary to the ASKAP EMU survey.

\subsection{Surveys with LOFAR}

\begin{figure} 
  \centering
  \includegraphics[scale=0.5]{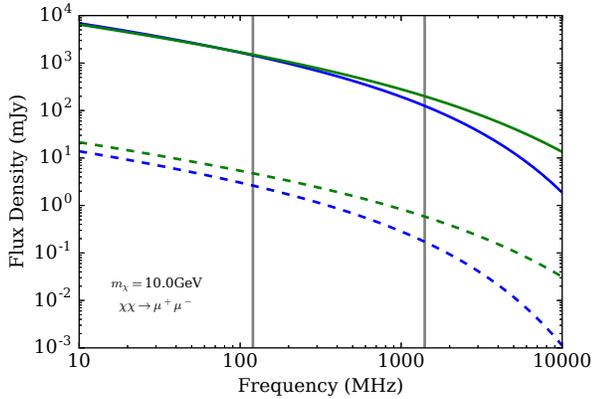}
  \caption{The expected flux density from dark matter annihilation as a function of frequency, for annihilation to muons, a dark matter mass of $m_{\chi}=10$~GeV and cross section $\langle \sigma v \rangle = 3\times10^{-26}$~cm$^3$s$^{-1}$. The solid blue line corresponds to $z=0.023$ and a ``non-cool-core'' magnetic field profile. The solid green line corresponds to $z=0.023$ and a ``cool-core'' magnetic field profile. The dashed blue line corresponds to $z=0.5$ and a ``non-cool-core'' magnetic field profile. The dashed green line corresponds to $z=0.5$ and a ``cool-core'' magnetic field profile. The vertical grey lines indicate the frequency of the LOFAR Tier 1 survey (120 MHz) and that of the EMU and APERTIF surveys (1.4 GHz) for reference.}
  \label{fig:signalfreq}
\end{figure}

The Low Frequency Array (LOFAR) is an international radio telescope array designed to operate in the $10-240$~MHz range \citep{vanHaarlem2013}. The bulk of the antennas are located throughout the Netherlands, with several international telescopes located throughout Europe. Several surveys are planned for LOFAR. The Multifrequency Snapshot Sky Survey (MSSS) is the first northern-sky survey from LOFAR and has already released preliminary results \citep{Heald2015}. The MSSS is designed to be a commissioning survey, reaching a depth of $10$~mJy at a resolution of $2$~arcmin over a wide range in frequency, from $30$ to $160$~MHz. 

The planned Tier 1 survey will survey the entire northern sky in the $15-180$~MHz band, and is supposed to achieve a depth of $\sim100~\mu$Jy at a $\sim5$~arcsec resolution at $120$~MHz \citep{vanHaarlem2013}. The Tier 1 survey will be highly complementary to the APERTIF surveys in particular, allowing for robust measurements of the spectral indices of the radio sources detected by both surveys over more than a decade in frequency.

LOFAR will also conduct deeper surveys over smaller areas.  For example the Tier 2 surveys will cover 550 deg$^2$ to a depth of $15~\mu$Jy at 150 MHz in a set of selected extragalactic, cluster, and supercluster fields.  Finally, a Tier 3 ultra-deep survey covering 83 deg$^2$ to a depth of $7~\mu$Jy at 150 MHz will be conducted and may be relevant to dark matter detection depending on if suitable target clusters fall within the survey region \citep{vanHaarlem2013}.

\section{Predictions for Dark Matter in Clusters}\label{sec:pred}

\subsection{Choice of Cluster Parameters}

\begin{figure} 
  \centering
  \includegraphics[scale=0.5]{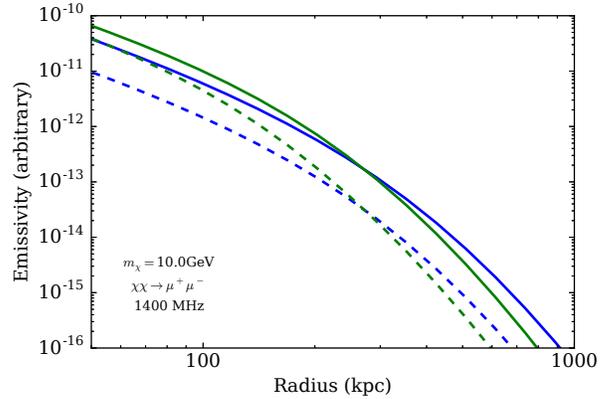}
  \caption{The expected, unnormalized emissivity from dark matter annihilation as a function of radius for annihilation to muons and a dark matter mass of $m_{\chi}=10$~GeV, at an observing frequency of $1400$~MHz. The solid blue line corresponds to $z=0.023$ and a ``non-cool-core'' magnetic field profile. The solid green line corresponds to $z=0.023$ and a ``cool-core'' magnetic field profile. The dashed blue line corresponds to $z=0.5$ and a ``non-cool-core'' magnetic field profile. The dashed green line corresponds to $z=0.5$ and a ``cool-core'' magnetic field profile.}
  \label{fig:signalrad}
\end{figure}

In order to constrain the dark matter annihilation cross section, we need to calculate the signal from dark matter and estimate the sensitivity of our chosen radio surveys to diffuse emission from a galaxy cluster. We choose to consider a dark matter halo mass of $10^{15}$~$M_{\odot}$, which is roughly the mass of the Coma Cluster. For this putative halo mass, we consider how the constraints on annihilation change over a redshift range of $0<z<0.5$ and two choices for the magnetic field profile (``cool-core'' and ``non-cool-core''; see Section~\ref{sec:magfield}). 

The synchrotron signal scales strongly with the observational frequency. Interestingly, the LOFAR surveys will access very low frequencies.  In Figure~\ref{fig:signalfreq}, we show how the flux density as calculated in Eq.~\ref{eq:ssynap} varies as a function of the observing frequency for a $10^{15}$~$M_{\odot}$ dark matter halo at two different redshifts, and for two different magnetic field models. Note that the drop in the magnitude of the signal for larger redshifts is due to the increasing distance. 

As is clear in Figure~\ref{fig:signalfreq}, the choice of observing frequency can have a dramatic effect on the estimated signal and therefore the constraints on dark matter annihilation. At lower frequencies, the differences between magnetic field models are minimized, especially for nearby objects. This is due to a complicated set of effects. First, the magnetic field profiles we use, a strong but sharply-peaked ``cool-core'' model and a weaker but flatter ``non-cool-core'' model, average out to similar values at $\sim300$~kpc, around our chosen ROI. Second, the dependence of the signal on the magnetic field scales with the strength of the field, the index $\alpha$ of the radio emission spectrum, and implicitly on the redshift via the effective magnetic field strength of the CMB. For a power-law radio emission spectrum, the dependence is as follows:

\begin{equation}\label{eq:Bdep}
  S_{\nu} \sim \nu^{-\alpha}\frac{B^{1+\alpha}}{B^2+B_{CMB,z=0}^2(1+z)^4}.
\end{equation}

Third, assuming a power-law spectrum, the radio spectral index $\alpha$ is related to the $e^{\pm}$ spectral index $p$: $\alpha = (p-1)/2$ (where $dN/dE \propto E^{-p}$). The shape of the $e^{\pm}$ spectrum depends on the dark matter mass and the annihilation channel. Generally, leptons and higher dark matter masses yield harder spectra than quarks and lower dark matter masses; $p$ typically varies from $2$ to $3$ (however, the actual spectra are more accurately described by power laws with cutoffs). Therefore, the dependence of the signal on the magnetic field is weaker for annihilation to muons than to quarks (see Figure~\ref{fig:mxsvpanel} and the discussion in Section~\ref{sec:DMlim}). Additionally, different observing frequencies essentially probe different parts of the underlying $e^{\pm}$ spectrum. At lower observing frequencies, the radio emission spectrum is harder than at higher observing frequencies, as seen in Figure~\ref{fig:signalfreq}. The combination of these effects account for the similarities in the signal for the different magnetic field profiles at low redshift and low observing frequencies. At high redshift, the effective CMB field starts to become relevant.

The signal, of course, also depends on the region of interest. We choose a region size that balances optimizing the signal-to-noise within the region, while also integrating over a physically meaningful area. In Figure~\ref{fig:signalrad}, we show how the emissivity, as calculated in Eq.~\ref{eq:jv}, varies as a function of radius. There is a turnover in the radial profile around the $\sim 100-300$~kpc range. We therefore choose a circular region with a fixed physical radius of $300$~kpc as the region in which we determine our signal and calculate the minimum detectable flux as discussed in the next section. This radius of $300$~kpc roughly corresponds to the core radius of the Coma Cluster. Somewhat interestingly, after $300$~kpc, the emissivities calculated with a ``cool-core'' magnetic field profile fall off more quickly than the ``non-cool-core''. If we were to choose a larger region size, the differences between these different magnetic field models would be smaller. However, this would also not appreciably boost the signal, while also increasing the noise, as we discuss in the next section.

\subsection{Estimation of Upper Limits on Diffuse Radio Emission}\label{sec:UL}

A conservative estimate for the values of the upper limits from the non-detection of a cluster in a radio survey can be derived by estimating the minimum detectable flux density integrated within some fixed region. The primary parameters we need in order to estimate upper limits on radio emission from future radio surveys are the size of the region of interest, the sensitivity of the radio survey, and the angular resolution of the survey. 

We use the quoted expected sensitivities for each survey in our calculations: for the ASKAP and APERTIF surveys, we assume a sensitivity of $10~\mu$Jy per beam at $1.4$~GHz, and for the LOFAR Tier 1 survey, we assume a sensitivity of $70~\mu$Jy per beam at $120$~MHz  \citep{2011PASA...28..215N, vanHaarlem2013,Shimwell2016}. While these surveys will have $\lesssim10$~arcsec resolution, we assume a gaussian beam with a FWHM size of $25$~arcsec~$\times$~$25$~arcsec. This larger beam size increases the sensitivity to extended emission, but is also small enough so that a circular area with a $300$~kpc radius at a redshift of $0.5$ is still sufficiently sampled by the beam ($300$~kpc corresponds to $49.2$~arcsec at $z=0.5$).

These sensitivities are close to the confusion limit for this resolution (see, e.g. \citealt{Condon2012} for a recent analysis of confusion noise in radio images, and references therein). However, when imaging diffuse emission, generally bright point sources are first removed using the full resolution data set ($5$~arcsec for LOFAR and $\lesssim10$~arcsec for ASKAP and APERTIF) and the remaining extended emission is imaged, which helps to mitigate issues of confusion noise from overlapping point sources in the restoring beam. In practice, the noise level of any image depends on the details of the imaging process and the quality of the data. We expect that the noise levels for different resolutions should be relatively similar for high quality, well-imaged maps. For example, pointed observations of diffuse emission with LOFAR already reach noise levels within a factor of $3$ of the assumed sensitivity above for a similar resolution \citep{VanWeeren2016}. Additionally, for a few clusters observed in the GMRT Radio Halo Survey \citep{Venturi2007,Venturi2008}, the noise levels are within a factor of $2-3$ of the confusion limit, after point-source subtraction and convolution with a larger beam (see also \citealt{Vernstrom2015}). The development of new  calibration and imaging techiques, such as facet calibration developed for LOFAR \citep{VanWeeren2016a}, will also help to improve the quality of images from upcoming surveys.

With these parameters, we estimate an upper limit on the diffuse radio emission detectable by a particular radio survey. We calculate the upper limit as the survey sensitivity per beam times the ratio of the region area over the beam area:

\begin{equation}\label{eq:UL}
  UL(r,z)=\sigma_{rms}\left( \frac{A_{ROI}(r,z)}{A_{beam}}\right)
\end{equation}

The area of the region of interest is:

\begin{equation}
  A_{ROI}(r,z)=\pi \theta_{ROI}(r,z)^2
\end{equation}

Here, $r$ is the radius of the region of interest, fixed to $300$~kpc. The angular radius $\theta_{ROI}$ is equal to this physical radius divided by the angular diameter distance at a given redshift.

The area of the beam is:

\begin{equation}
  A_{beam}=\frac{\pi}{4ln(2)}\theta_{beam}^2
\end{equation}

Where $\theta_{beam}$ is the usual quoted FWHM value for a gaussian beam, assumed to be $25$~arcsec in our calculations.

This methodology for estimating upper limits is similar to that in \citet[e.g.,][]{Cassano2009,Cassano2012}, where they estimate the minimum detectable flux density from a radio halo to determine the number of halos that are potentially detectable by upcoming radio surveys.

\subsection{Constraints on Dark Matter Annihilation}\label{sec:DMlim}
\begin{figure*}
\gridline{\fig{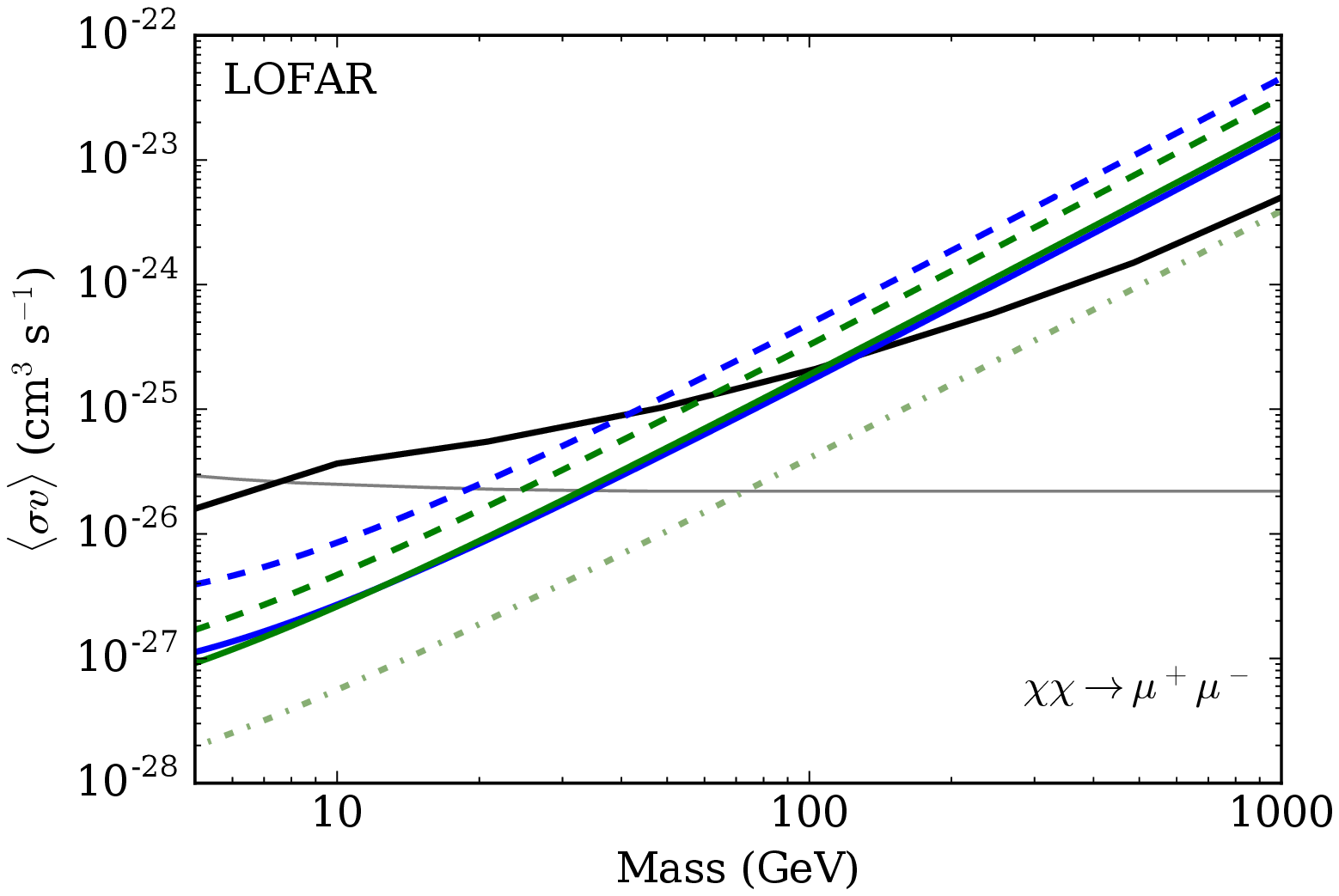}{0.45\textwidth}{(a)}
          \fig{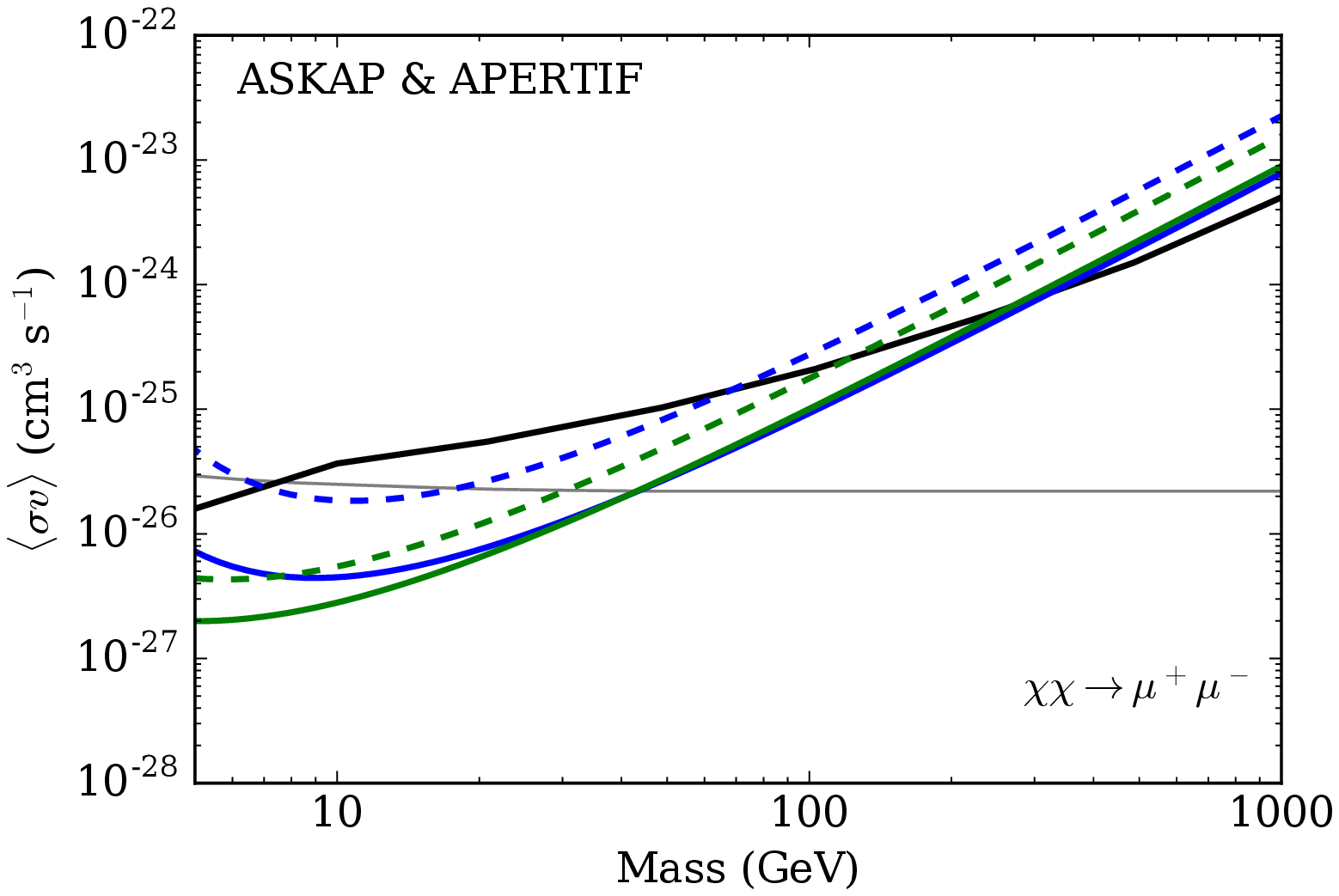}{0.45\textwidth}{(b)}}
\gridline{\fig{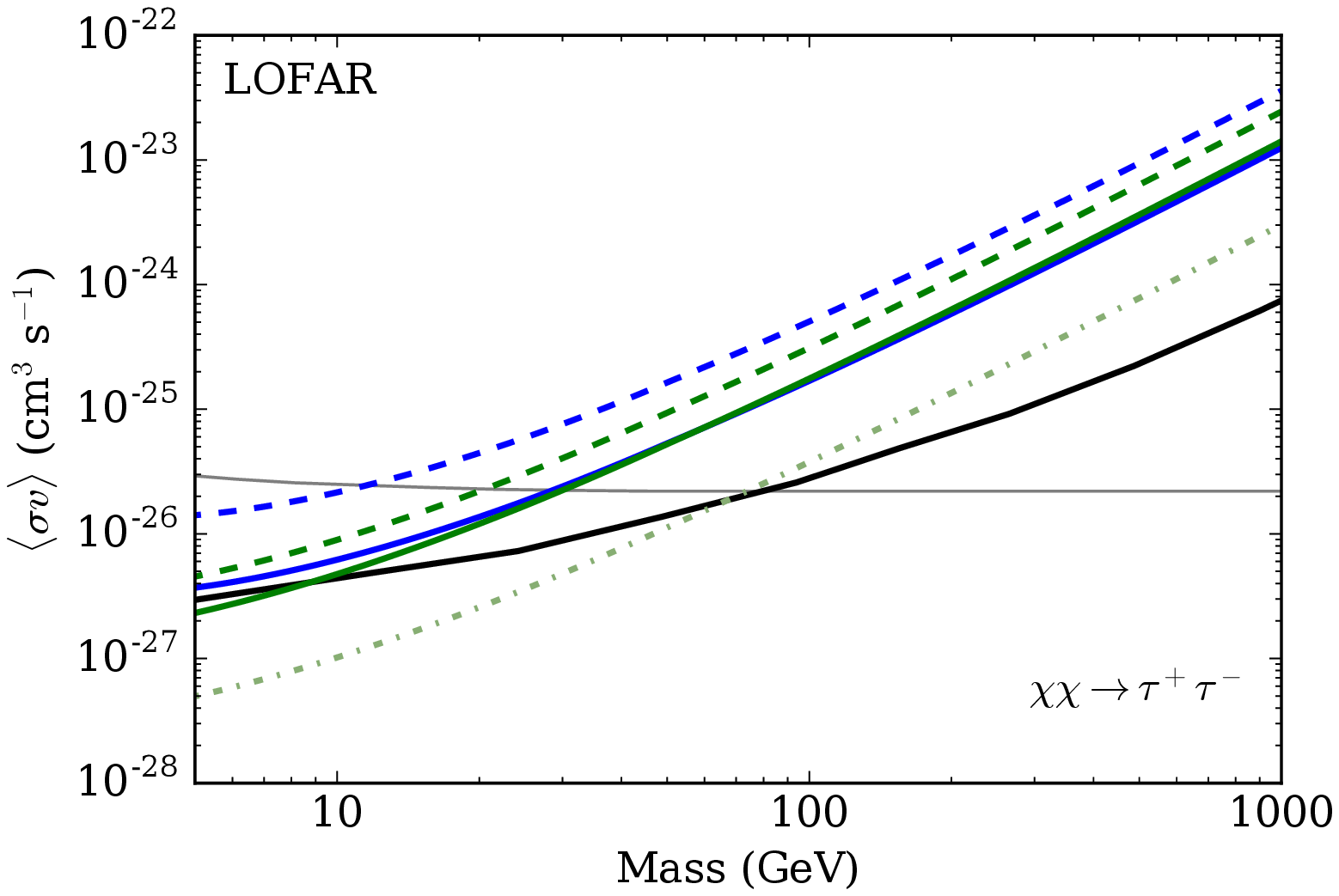}{0.45\textwidth}{(d)}
          \fig{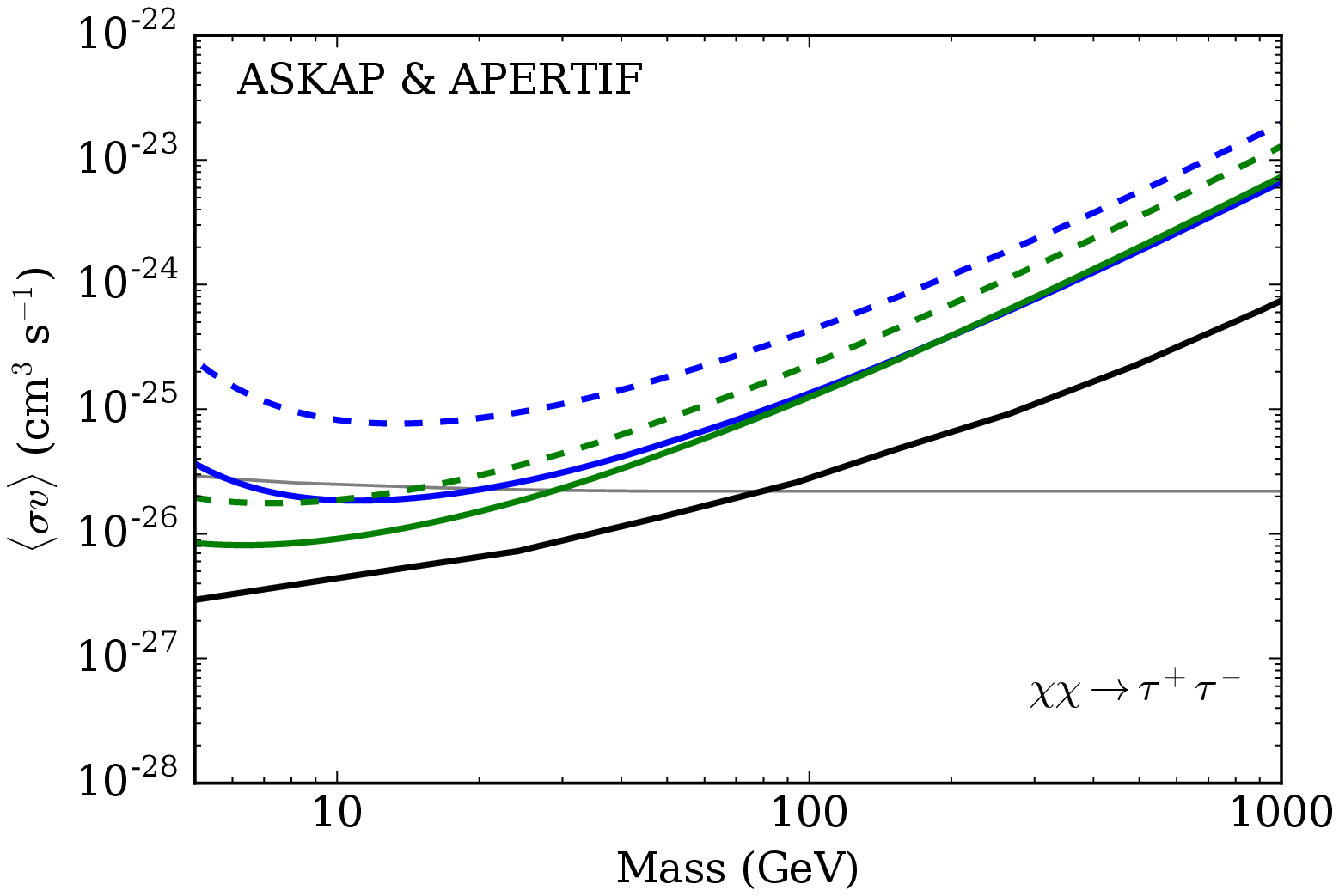}{0.45\textwidth}{(e)}}
\gridline{\fig{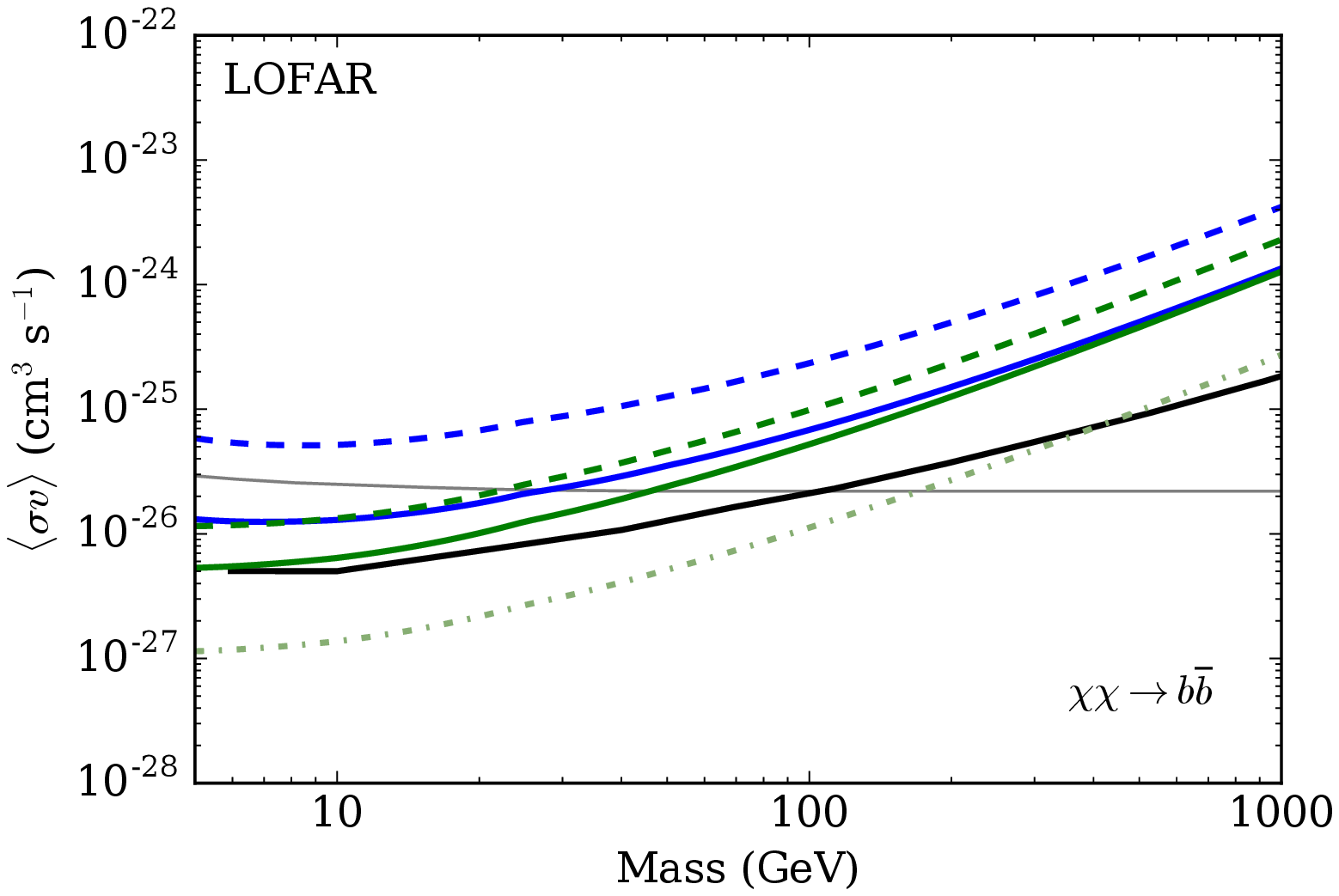}{0.45\textwidth}{(d)}
          \fig{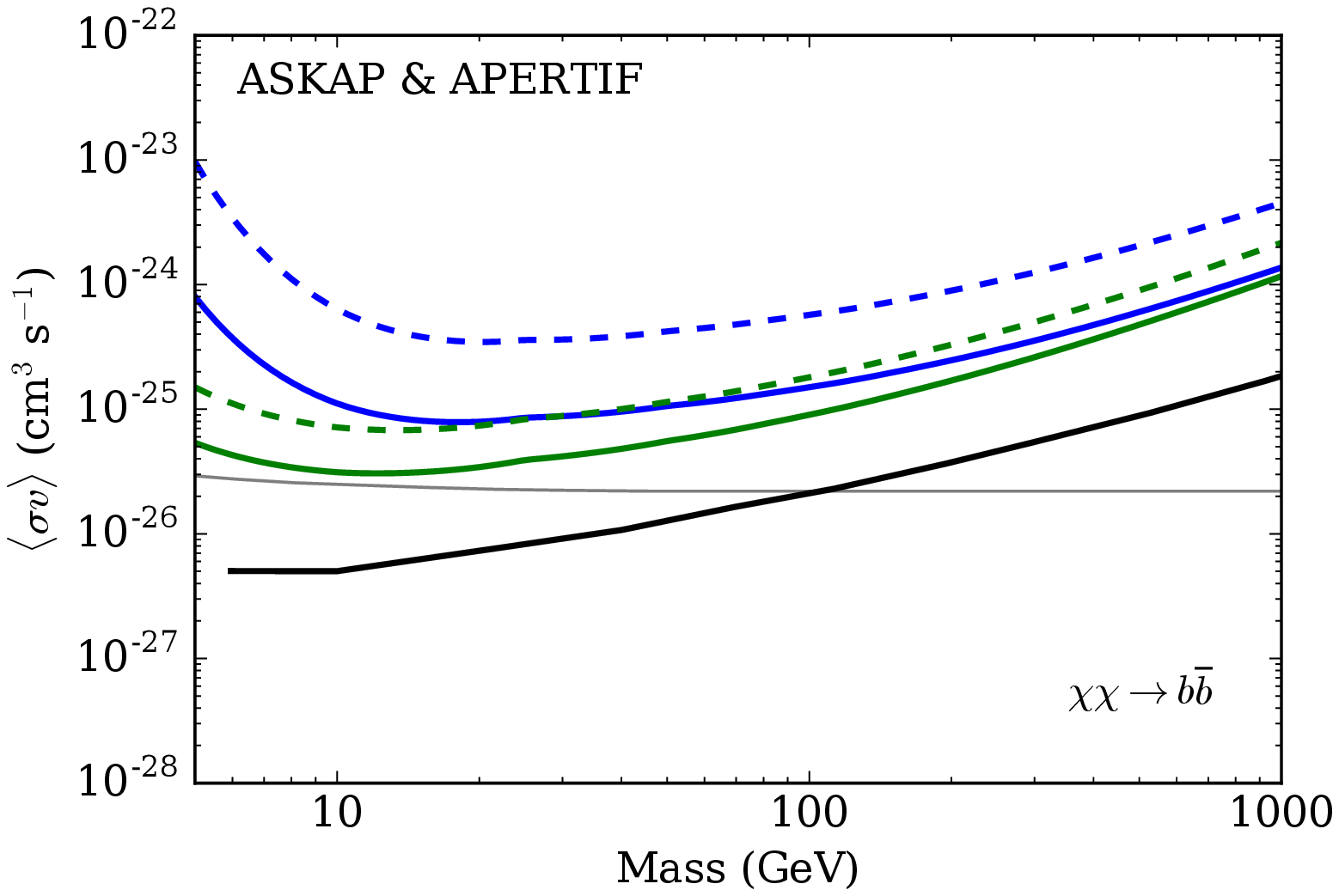}{0.45\textwidth}{(e)}}
\caption{Constraints on the annihilation cross section as a function of dark matter mass for annihilation to muons (a,b), tau leptons (c,d) and a $b\bar{b}$ quark pair. In all figures, these limits assume a beam size of $25$~arcsec. Lines for all figures are as follows: $z=0.023$ with a ``non-cool-core'' magnetic field profile (solid blue), $z=0.023$ with a ``cool-core'' magnetic field profile (solid green), $z=0.5$ and ``non-cool-core'' (dashed blue), and $z=0.5$ and ``cool-core'' (dashed green). The black line represents the constraints from the non-detection of gamma rays in dwarf galaxies, from Figure 8 in \citet{Ackermann2015}. The gray line indicates the benchmark ``thermal relic'' cross-section \citep{2012PhRvD..86b3506S}. The left-hand column (a,c,e) of 3 figures show limits from LOFAR for the labeled annihilation channels, assuming a sensitivity of  $70~\mu$Jy per beam for the Tier 1 survey. The right-hand column (b,d,f) show limits from ASKAP or APERTIF surveys at $1400$~MHz for the labeled annihilation channels, assuming a sensitivity of $10~\mu$Jy per beam.}
\label{fig:mxsvpanel}
\end{figure*}

In the multipanel Figure~\ref{fig:mxsvpanel}, we show the upper limits on the predicted annihilation cross section as a function of dark matter mass for different magnetic field profiles, redshifts, and annihilation channels. Since the synchrotron signal from dark matter annihilation depends on the resulting electron and positron spectra, the strongest constraints result from annihilation to muons. This channel produces, from muon decay, a much harder spectrum of electrons and positrons compared to channels where electron-positron production stems from hadronization and charged-pion decay. Although it is not shown in the plots, the constraint curves at $120$~MHz do turn over at about $\sim2$~GeV. (The turnover in the constraint curves at $1400$~MHz is seen in the figures clearly, at around $10$~GeV.)

At energies $\lesssim 30$~GeV, the predicted upper limits on the annihilation cross section are $\gtrsim 1$ order of magnitude below the thermal relic cross section.\footnote{The thermal relic cross section is often quoted as $3\times10^{-26}$~cm$^3$s$^{-1}$. We instead show in the figures the more accurate cross section calculated by \citet{2012PhRvD..86b3506S}, which includes explicitly the WIMP mass dependence and uses the best-fit cosmological parameters from recent observations to derive a cross section that is accurate at the level of a few percent.}

\begin{figure} 
  \centering
  \includegraphics[scale=0.5]{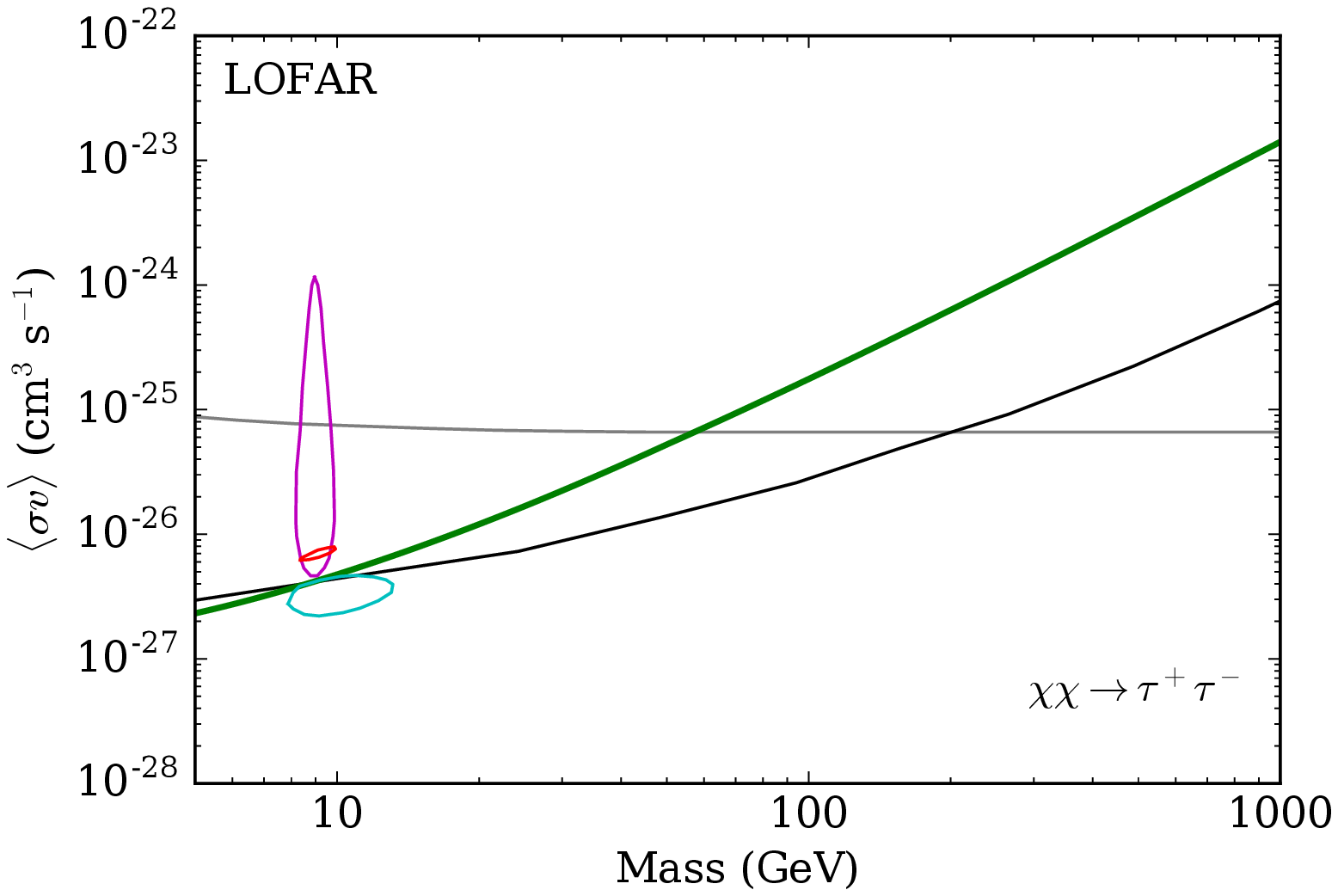}
  \caption{Comparison between predicted limits with radio surveys and best-fit models to the Galactic Center GeV excess for annihilation to taus. Closed contours are $95\%$ confidence limits from the following studies: magenta: \citet{Abazajian2016}, red: \citet{Daylan2016}, cyan: \citet{Calore2015a}. The green lines represent limits for the LOFAR Tier 1 survey, at $120$~MHz, sensitivity of $70~\mu$Jy per beam with a $25$~arcsec beam. The solid green line corresponds to $z=0.023$ and a ``cool-core'' magnetic field profile for Tier 1. Note that a non-cool-core magnetic field model gives almost identical constraints as shown in Figures~\ref{fig:mxsvpanel}(b,c).  The black line represents the constraints from the non-detection of gamma rays in dwarf galaxies, from Figure 8 in \citet{Ackermann2015}. The gray horizontal line indicates the benchmark ``thermal relic'' cross-section \citep{2012PhRvD..86b3506S}.}
  \label{fig:mxsvGCE}
\end{figure}

In each plot of the annihilation cross section limits, we also compare our predicted limits to those from the combined non-detection of gamma rays from a sample of 15 dwarf galaxies with \textit{Fermi}-LAT, which are the best single-channel cross section limits to date \citep{Ackermann2015}. In the case of annihilation to muons, our predicted limits are up to an order of magnitude more sensitive than the limits from dwarf galaxies even for the shallow LOFAR Tier 1 and ASKAP EMU surveys. For other annihilation channels, the limits from radio are generally less sensitive than the gamma ray dwarf limits for the shallow, wide-area surveys, because annihilation to tau leptons and to bottom quarks yield many gamma rays, but relatively fewer electrons and positrons. In Figure~\ref{fig:mxsvGCE}, we show that the predicted limits from LOFAR nondetections will rule out some of the the best-fit models to the Galactic Center Excess \citep{Abazajian2016,Daylan2016,Calore2015a}. Similarly, deep pointed observations of appropriate clusters with LOFAR, ASKAP, or APERTIF can yield the strongest indirect constraints on a large range of dark matter annihilation models.

The shape of the predicted constraint curves vary depending on the observing frequency for a given dark matter particle mass, since a different part of the underlying electron/positron spectrum is being probed. At low dark matter masses, the non-detection of diffuse emission from clusters yield the most constraining limits at lower observing frequencies, in the range of the LOFAR surveys, while at higher dark matter masses, the higher observing frequencies of the EMU and APERTIF surveys yield better limits. Notably, at $120$~MHz, the effect of the magnetic field profile on the limits is smaller than at $1.4$~GHz and virtually negligible at high dark matter masses for all annihilation channels. This makes LOFAR a better instrument to constrain dark matter from clusters, since the magnetic field profiles and strengths in clusters are somewhat uncertain.

Finally, the effect of redshift on the cross-section limits is observed to be less severe than the variation in the signal from Figure~\ref{fig:signalfreq} would imply. This is because while the synchrotron flux decreases as a function of distance and therefore redshift, the minimum detectable flux for a given survey also decreases with redshift for a fixed physical region of interest; the smaller angular size at higher redshift leads to a lower background in the cluster region.  The limits from a cluster at $z=0.5$ are at most an order of magnitude larger than those at $z=0.023$ (the redshift of the Coma Cluster), and are only a factor of $\sim3$ larger in most cases. This implies that limits from radio emission are less sensitive to the distance of the object, which allows for a larger sample to be analyzed, in comparison to, for example, indirect detection studies in the gamma-ray band.

\section{Conclusions}\label{sec:end}

In this paper, we estimate the synchrotron signal from dark matter annihilation in a galaxy cluster, and make predictions for constraints on dark matter annihilation from the non-detection of radio emission from clusters with upcoming radio surveys. We explore which factors are most important in determining a sample for future indirect detection studies using the non-detection of radio emission from galaxy clusters, including the optimal redshift range, region of interest, frequency of observation, and magnetic field profile.

In summary, we find that the synchrotron signal as a function of radius starts to fall off in the $100-300$~kpc range for all redshift ranges and different magnetic field configurations; this is approximately the size of the core radius of the cluster (and the characteristic radius of the dark matter halo). If the upper limit on the minimum detectable integrated flux for a given survey increases linearly with area, as we have assumed in this paper, then this region of interest of $\sim300$~kpc used to determine upper limits will maximize the signal-to-noise.  Over this region of interest, the differences between magnetic field profiles are also minimal.

We find that increasing the redshift does not have a strong effect on the predicted limits for the annihilation cross section. This is not necessarily surprising, as we use a brightness threshold to derive upper limits on the predicted radio emission. Therefore, there is a larger pool of clusters from which to select a sample for future indirect detection studies in the radio band, than in, e.g., the gamma-ray band, where the best constraints result from samples of local objects.

The best constraints on dark matter annihilation from indirect detection to-date are those from the non-detection of gamma-rays from dwarf galaxies with \textit{Fermi} \citep{Ackermann2015}. However, future radio surveys have the potential to surpass those constraints, by an order of magnitude or more for dark matter masses $\lesssim 100$~GeV and for annihilation to muons. For annihilation to other channels, the limits from dwarfs remain competitive with those predicted for the shallow, wide-area ASKAP, APERTIF, and the LOFAR Tier 1 surveys.

Deeper planned surveys can potentially yield even more stringent upper limits on radio emission from clusters, and thus tighter constraints on dark matter models. For example, the LOFAR Tier 2 survey is expected to be approximately 5 times deeper than the Tier 1 survey \citep{vanHaarlem2013}, so the upper limits on dark matter annihilation could be, correspondingly, up to 5 times deeper. Further into the future, anticipated surveys from SKA will be deeper than AKSAP/APERTIF and LOFAR by a factor of 3 or more \citep[][e.g.,]{Prandoni2014}. Similarly, pointed observations of well-selected target clusters with the instruments studied here would yield novel indirect constraints on dark matter models.

Future radio observations might then reveal the first signal from dark matter annihilation.  The difficulty in this case would be to disentangle a dark matter signal from astrophysical emission like radio halos. The fact that the dark matter signal is expected to be similar for clusters over a large redshift range and is not highly magnetic-field dependent at some frequencies gives a method of testing the origin of a potential dark matter signature.

\acknowledgments
We thank Christoph Weniger for useful discussions and a careful reading of the draft. E.S. acknowledges support from the Cota-Robles Fellowship. This material is based upon work supported by the National Science Foundation under Grant No. 1517545. SP is partly supported by the U.S. Department of Energy under Grant No. DE-SC0010107. 

\bibliography{PredRadioDM}

\begin{thebibliography}{}
\expandafter\ifx\csname natexlab\endcsname\relax\def\natexlab#1{#1}\fi

\bibitem[{Abazajian \& Keeley(2016)}]{Abazajian2016}
Abazajian, K.~N., \& Keeley, R.~E. 2016, Physical Review D, 93, 083514

\bibitem[{Ackermann {et~al.}(2015)Ackermann, Albert, Anderson, Atwood, Baldini,
  Barbiellini, Bastieri, Bechtol, Bellazzini, Bissaldi, Blandford, Bloom,
  Bonino, Bottacini, Brandt, Bregeon, Bruel, Buehler, Caliandro, Cameron,
  Caputo, Caragiulo, Caraveo, Cecchi, Charles, Chekhtman, Chiang, Chiaro,
  Ciprini, Claus, Cohen-Tanugi, Conrad, Cuoco, Cutini, D'Ammando, {De Angelis},
  {De Palma}, Desiante, Digel, {Di Venere}, Drell, Drlica-Wagner, Essig,
  Favuzzi, Fegan, Ferrara, Focke, Franckowiak, Fukazawa, Funk, Fusco, Gargano,
  Gasparrini, Giglietto, Giordano, Giroletti, Glanzman, Godfrey, Gomez-Vargas,
  Grenier, Guiriec, Gustafsson, Hays, Hewitt, Horan, Jogler, J{\'{o}}hannesson,
  Kuss, Larsson, Latronico, Li, Li, {Llena Garde}, Longo, Loparco, Lubrano,
  Malyshev, Mayer, Mazziotta, McEnery, Meyer, Michelson, Mizuno, Moiseev,
  Monzani, Morselli, Murgia, Nuss, Ohsugi, Orienti, Orlando, Ormes, Paneque,
  Perkins, Pesce-Rollins, Piron, Pivato, Porter, Rain{\`{o}}, Rando, Razzano,
  Reimer, Reimer, Ritz, S{\'{a}}nchez-Conde, Schulz, Sehgal, Sgr{\`{o}},
  Siskind, Spada, Spandre, Spinelli, Strigari, Tajima, Takahashi, Thayer,
  Tibaldo, Torres, Troja, Vianello, Werner, Winer, Wood, Wood, Zaharijas, \&
  Zimmer}]{Ackermann2015}
Ackermann, M., Albert, A., Anderson, B., {et~al.} 2015, Physical Review
  Letters, 115, 1

\bibitem[{{Bergstr{\"o}m}(2000)}]{2000RPPh...63..793B}
{Bergstr{\"o}m}, L. 2000, Reports on Progress in Physics, 63, 793

\bibitem[{{Bertone} {et~al.}(2005){Bertone}, {Hooper}, \&
  {Silk}}]{2005PhR...405..279B}
{Bertone}, G., {Hooper}, D., \& {Silk}, J. 2005, \physrep, 405, 279

\bibitem[{Bonafede {et~al.}(2010)Bonafede, Feretti, Murgia, Govoni, Giovannini,
  Dallacasa, Dolag, \& Taylor}]{Bonafede2010}
Bonafede, A., Feretti, L., Murgia, M., {et~al.} 2010, Astronomy and
  Astrophysics, 513, A30

\bibitem[{Briel {et~al.}(1992)Briel, Henry, \& Bohringer}]{Briel1992}
Briel, U.~G., Henry, J.~P., \& Bohringer, H. 1992, Astronomy {\&} Astrophysics,
  259, L31

\bibitem[{{Brunetti} {et~al.}(2012){Brunetti}, {Blasi}, {Reimer}, {Rudnick},
  {Bonafede}, \& {Brown}}]{2012MNRAS.426..956B}
{Brunetti}, G., {Blasi}, P., {Reimer}, O., {et~al.} 2012, \mnras, 426, 956

\bibitem[{{Brunetti} \& {Jones}(2014)}]{2014IJMPD..2330007B}
{Brunetti}, G., \& {Jones}, T.~W. 2014, International Journal of Modern Physics
  D, 23, 1430007

\bibitem[{{Brunetti} \& {Lazarian}(2016)}]{2016MNRAS.458.2584B}
{Brunetti}, G., \& {Lazarian}, A. 2016, \mnras, 458, 2584

\bibitem[{Buote {et~al.}(2007)Buote, Gastaldello, Humphrey, Zappacosta,
  Bullock, Brighenti, \& Mathews}]{Buote2007}
Buote, D.~A., Gastaldello, F., Humphrey, P.~J., {et~al.} 2007, The
  Astrophysical Journal, 664, 123

\bibitem[{Calore {et~al.}(2015)Calore, Cholis, \& Weniger}]{Calore2015a}
Calore, F., Cholis, I., \& Weniger, C. 2015, Journal of Cosmology and
  Astroparticle Physics, 2015, 038

\bibitem[{Carilli \& Taylor(2002)}]{Carilli2002}
Carilli, C.~L., \& Taylor, G.~B. 2002, Annual Review of Astronomy and
  Astrophysics, 40, 319

\bibitem[{Cassano {et~al.}(2012)Cassano, Brunetti, Norris, R{\"{o}}ttgering,
  Johnston-Hollitt, \& Trasatti}]{Cassano2012}
Cassano, R., Brunetti, G., Norris, R.~P., {et~al.} 2012, Astronomy {\&}
  Astrophysics, 548, A100

\bibitem[{Cassano {et~al.}(2009)Cassano, Brunetti, R{\"{o}}ttgering, \&
  Br{\"{u}}ggen}]{Cassano2009}
Cassano, R., Brunetti, G., R{\"{o}}ttgering, H. J.~A., \& Br{\"{u}}ggen, M.
  2009, Astronomy and Astrophysics, 509, A68

\bibitem[{{Cassano} {et~al.}(2010){Cassano}, {Ettori}, {Giacintucci},
  {Brunetti}, {Markevitch}, {Venturi}, \& {Gitti}}]{2010ApJ...721L..82C}
{Cassano}, R., {Ettori}, S., {Giacintucci}, S., {et~al.} 2010, \apjl, 721, L82

\bibitem[{Charles {et~al.}(2016)Charles, Sanchez-Conde, Anderson, Caputo,
  Cuoco, {Di Mauro}, Drlica-Wagner, Gomez-Vargas, Meyer, Tibaldo, Wood,
  Zaharijas, Zimmer, Ajello, Albert, Baldini, Bechtol, Bloom, Ceraudo,
  Cohen-Tanugi, Digel, Gaskins, Gustafsson, Mirabal, \& Razzano}]{Charles2016}
Charles, E., Sanchez-Conde, M., Anderson, B., {et~al.} 2016, 1

\bibitem[{Chen {et~al.}(2007)Chen, Reiprich, B{\"{o}}hringer, Ikebe, \&
  Zhang}]{Chen2007}
Chen, Y., Reiprich, T.~H., B{\"{o}}hringer, H., Ikebe, Y., \& Zhang, Y.-Y.
  2007, Astronomy and Astrophysics, 466, 805

\bibitem[{Clowe {et~al.}(2006)Clowe, Bradac, Gonzalez, Markevitch, Randall,
  Jones, \& Zaritsky}]{Clowe2006}
Clowe, D., Bradac, M., Gonzalez, A.~H., {et~al.} 2006, The Astrophysical
  Journal, 648, L109

\bibitem[{Colafrancesco {et~al.}(2006)Colafrancesco, Profumo, \&
  Ullio}]{Colafrancesco2006}
Colafrancesco, S., Profumo, S., \& Ullio, P. 2006, Astronomy and Astrophysics,
  455, 21

\bibitem[{Condon {et~al.}(2012)Condon, Cotton, Fomalont, Kellermann, Miller,
  Perley, Scott, Vernstrom, \& Wall}]{Condon2012}
Condon, J.~J., Cotton, W.~D., Fomalont, E.~B., {et~al.} 2012, The Astrophysical
  Journal, 758, 23

\bibitem[{Conrad {et~al.}(2015)Conrad, Cohen-Tanugi, \& Strigari}]{Conrad2015}
Conrad, J., Cohen-Tanugi, J., \& Strigari, L.~E. 2015, Journal of Experimental
  and Theoretical Physics, 121, 1104

\bibitem[{Daylan {et~al.}(2016)Daylan, Finkbeiner, Hooper, Linden, Portillo,
  Rodd, \& Slatyer}]{Daylan2016}
Daylan, T., Finkbeiner, D.~P., Hooper, D., {et~al.} 2016, Physics of the Dark
  Universe, 12, 1

\bibitem[{Feng(2010)}]{Feng2010}
Feng, J.~L. 2010, Annual Review of Astronomy and Astrophysics, 48, 495

\bibitem[{{Feretti} {et~al.}(2012){Feretti}, {Giovannini}, {Govoni}, \&
  {Murgia}}]{2012A&ARv..20...54F}
{Feretti}, L., {Giovannini}, G., {Govoni}, F., \& {Murgia}, M. 2012, \aapr, 20,
  54

\bibitem[{Gaskins(2016)}]{Gaskins2016}
Gaskins, J.~M. 2016, Contemporary Physics, 57, 496

\bibitem[{Gondolo {et~al.}(2004)Gondolo, Edsj{\"{o}}, Ullio, Bergstr{\"{o}}m,
  Schelke, \& Baltz}]{Gondolo2004}
Gondolo, P., Edsj{\"{o}}, J., Ullio, P., {et~al.} 2004, Journal of Cosmology
  and Astroparticle Physics, 2004, 008

\bibitem[{Heald {et~al.}(2015)Heald, Pizzo, Orr{\'{u}}, Breton, Carbone,
  Ferrari, Hardcastle, Jurusik, Macario, Mulcahy, Rafferty, Asgekar, Brentjens,
  Fallows, Frieswijk, Toribio, Adebahr, Arts, Bell, Bonafede, Bray, Broderick,
  Cantwell, Carroll, Cendes, Clarke, Croston, Daiboo, de~Gasperin, Gregson,
  Harwood, Hassall, Heesen, Horneffer, van~der Horst, Iacobelli, Jeli{\'{c}},
  Jones, Kant, Kokotanekov, Martin, McKean, Morabito, Nikiel-Wroczyński,
  Offringa, Pandey, Pandey-Pommier, Pietka, Pratley, Riseley, Rowlinson,
  Sabater, Scaife, Scheers, Sendlinger, Shulevski, Sipior, Sobey, Stewart,
  Stroe, Swinbank, Tasse, Tr{\"{u}}stedt, Varenius, van Velzen, Vilchez, van
  Weeren, Wijnholds, Williams, de~Bruyn, Nijboer, Wise, Alexov, Anderson,
  Avruch, Beck, Bell, van Bemmel, Bentum, Bernardi, Best, Breitling, Brouw,
  Br{\"{u}}ggen, Butcher, Ciardi, Conway, de~Geus, de~Jong, de~Vos, Deller,
  Dettmar, Duscha, Eisl{\"{o}}ffel, Engels, Falcke, Fender, Garrett,
  Grie{\ss}meier, Gunst, Hamaker, Hessels, Hoeft, H{\"{o}}randel, Holties,
  Intema, Jackson, J{\"{u}}tte, Karastergiou, Klijn, Kondratiev, Koopmans,
  Kuniyoshi, Kuper, Law, van Leeuwen, Loose, Maat, Markoff, McFadden,
  McKay-Bukowski, Mevius, Miller-Jones, Morganti, Munk, Nelles, Noordam,
  Norden, Paas, Polatidis, Reich, Renting, R{\"{o}}ttgering, Schoenmakers,
  Schwarz, Sluman, Smirnov, Stappers, Steinmetz, Tagger, Tang, ter Veen,
  Thoudam, Vermeulen, Vocks, Vogt, Wijers, Wucknitz, Yatawatta, \&
  Zarka}]{Heald2015}
Heald, G.~H., Pizzo, R.~F., Orr{\'{u}}, E., {et~al.} 2015, Astronomy {\&}
  Astrophysics, 582, A123

\bibitem[{Hu \& Kravtsov(2003)}]{Hu2003}
Hu, W., \& Kravtsov, A.~V. 2003, The Astrophysical Journal, 584, 702

\bibitem[{{Jeltema} \& {Profumo}(2011)}]{2011ApJ...728...53J}
{Jeltema}, T.~E., \& {Profumo}, S. 2011, \apj, 728, 53

\bibitem[{{Jeltema} \& {Profumo}(2012)}]{2012MNRAS.421.1215J}
---. 2012, \mnras, 421, 1215

\bibitem[{Johnston {et~al.}(2008)Johnston, Taylor, Bailes, Bartel, Baugh,
  Bietenholz, Blake, Braun, Brown, Chatterjee, Darling, Deller, Dodson,
  Edwards, Ekers, Ellingsen, Feain, Gaensler, Haverkorn, Hobbs, Hopkins,
  Jackson, James, Joncas, Kaspi, Kilborn, Koribalski, Kothes, Landecker, Lenc,
  Lovell, Macquart, Manchester, Matthews, McClure-Griffiths, Norris, Pen,
  Phillips, Power, Protheroe, Sadler, Schmidt, Stairs, Staveley-Smith, Stil,
  Tingay, Tzioumis, Walker, Wall, \& Wolleben}]{Johnston2008}
Johnston, S., Taylor, R., Bailes, M., {et~al.} 2008, Experimental Astronomy,
  22, 151

\bibitem[{{Jungman} {et~al.}(1996){Jungman}, {Kamionkowski}, \&
  {Griest}}]{1996PhR...267..195J}
{Jungman}, G., {Kamionkowski}, M., \& {Griest}, K. 1996, \physrep, 267, 195

\bibitem[{Lokas \& Mamon(2003)}]{Lokas2003}
Lokas, E.~L., \& Mamon, G.~A. 2003, Monthly Notices of the Royal Astronomical
  Society, 343, 401

\bibitem[{Murgia {et~al.}(2004)Murgia, Govoni, Feretti, Giovannini, Dallacasa,
  Fanti, Taylor, \& Dolag}]{Murgia2004}
Murgia, M., Govoni, F., Feretti, L., {et~al.} 2004, Astronomy and Astrophysics,
  424, 429

\bibitem[{Navarro {et~al.}(1996)Navarro, Frenk, \& White}]{Navarro1996}
Navarro, J.~F., Frenk, C.~S., \& White, S. D.~M. 1996, The Astrophysical
  Journal, 462, 563

\bibitem[{Navarro {et~al.}(1997)Navarro, Frenk, \& White}]{Navarro1997}
---. 1997, The Astrophysical Journal, 490, 493

\bibitem[{{Norris} {et~al.}(2011){Norris}, {Hopkins}, {Afonso}, {Brown},
  {Condon}, {Dunne}, {Feain}, {Hollow}, {Jarvis}, {Johnston-Hollitt}, {Lenc},
  {Middelberg}, {Padovani}, {Prandoni}, {Rudnick}, {Seymour}, {Umana},
  {Andernach}, {Alexander}, {Appleton}, {Bacon}, {Banfield}, {Becker}, {Brown},
  {Ciliegi}, {Jackson}, {Eales}, {Edge}, {Gaensler}, {Giovannini}, {Hales},
  {Hancock}, {Huynh}, {Ibar}, {Ivison}, {Kennicutt}, {Kimball}, {Koekemoer},
  {Koribalski}, {L{\'o}pez-S{\'a}nchez}, {Mao}, {Murphy}, {Messias},
  {Pimbblet}, {Raccanelli}, {Randall}, {Reiprich}, {Roseboom},
  {R{\"o}ttgering}, {Saikia}, {Sharp}, {Slee}, {Smail}, {Thompson}, {Urquhart},
  {Wall}, \& {Zhao}}]{2011PASA...28..215N}
{Norris}, R.~P., {Hopkins}, A.~M., {Afonso}, J., {et~al.} 2011, \pasa, 28, 215

\bibitem[{Porter {et~al.}(2011)Porter, Johnson, \& Graham}]{Porter2011}
Porter, T.~a., Johnson, R.~P., \& Graham, P.~W. 2011, Annual Review of
  Astronomy and Astrophysics, 49, 155

\bibitem[{Prandoni \& Seymour(2014)}]{Prandoni2014}
Prandoni, I., \& Seymour, N. 2014, Proceedings of Science, 9-13-June-2014, 1

\bibitem[{Profumo(2013)}]{Profumo:2013yn}
Profumo, S. 2013, in {Proceedings, Theoretical Advanced Study Institute in
  Elementary Particle Physics: Searching for New Physics at Small and Large
  Scales (TASI 2012)}, 143--189

\bibitem[{Profumo \& Ullio(2010)}]{Profumo:2010ya}
Profumo, S., \& Ullio, P. 2010, arXiv:1001.4086

\bibitem[{Shimwell {et~al.}(2016)Shimwell, R{\"{o}}ttgering, Best, Williams,
  Dijkema, {De Gasperin}, Hardcastle, Heald, Hoang, Horneffer, Intema, Mahony,
  Mandal, Mechev, Morabito, Oonk, Rafferty, Retana-Montenegro, Sabater, \&
  Tasse}]{Shimwell2016}
Shimwell, T.~W., R{\"{o}}ttgering, H. J.~A., Best, P.~N., {et~al.} 2016, J. P.
  McKean D. J. McKay N. R. Mohan R. F. Pizzo I. Prandoni D. J. Schwarz, 6,
  arXiv:1611.02700

\bibitem[{{Steigman} {et~al.}(2012){Steigman}, {Dasgupta}, \&
  {Beacom}}]{2012PhRvD..86b3506S}
{Steigman}, G., {Dasgupta}, B., \& {Beacom}, J.~F. 2012, \prd, 86, 023506

\bibitem[{Storm {et~al.}(2013)Storm, Jeltema, Profumo, \& Rudnick}]{Storm2013}
Storm, E., Jeltema, T.~E., Profumo, S., \& Rudnick, L. 2013, The Astrophysical
  Journal, 768, 106

\bibitem[{{Storm} {et~al.}(2015){Storm}, {Jeltema}, \&
  {Rudnick}}]{2015MNRAS.448.2495S}
{Storm}, E., {Jeltema}, T.~E., \& {Rudnick}, L. 2015, \mnras, 448, 2495

\bibitem[{Taylor {et~al.}(2006)Taylor, Gugliucci, Fabian, Sanders, Gentile, \&
  Allen}]{Taylor2006}
Taylor, G.~B., Gugliucci, N.~E., Fabian, a.~C., {et~al.} 2006, Monthly Notices
  of the Royal Astronomical Society, 368, 1500

\bibitem[{Vacca {et~al.}(2012)Vacca, Murgia, Govoni, Feretti, Giovannini,
  Perley, \& Taylor}]{Vacca2012}
Vacca, V., Murgia, M., Govoni, F., {et~al.} 2012, Astronomy {\&} Astrophysics,
  540, A38

\bibitem[{van Haarlem {et~al.}(2013)van Haarlem, Wise, Gunst, Heald, McKean,
  Hessels, de~Bruyn, Nijboer, Swinbank, Fallows, Brentjens, Nelles, Beck,
  Falcke, Fender, H{\"{o}}randel, Koopmans, Mann, Miley, R{\"{o}}ttgering,
  Stappers, Wijers, Zaroubi, van~den Akker, Alexov, Anderson, Anderson, van
  Ardenne, Arts, Asgekar, Avruch, Batejat, B{\"{a}}hren, Bell, Bell, van
  Bemmel, Bennema, Bentum, Bernardi, Best, B{\^{\i}}rzan, Bonafede, Boonstra,
  Braun, Bregman, Breitling, van~de Brink, Broderick, Broekema, Brouw,
  Br{\"{u}}ggen, Butcher, van Cappellen, Ciardi, Coenen, Conway, Coolen,
  Corstanje, Damstra, Davies, Deller, Dettmar, van Diepen, Dijkstra, Donker,
  Doorduin, Dromer, Drost, van Duin, Eisl{\"{o}}ffel, van Enst, Ferrari,
  Frieswijk, Gankema, Garrett, de~Gasperin, Gerbers, de~Geus, Grie{\ss}meier,
  Grit, Gruppen, Hamaker, Hassall, Hoeft, Holties, Horneffer, van~der Horst,
  van Houwelingen, Huijgen, Iacobelli, Intema, Jackson, Jelic, de~Jong, Juette,
  Kant, Karastergiou, Koers, Kollen, Kondratiev, Kooistra, Koopman, Koster,
  Kuniyoshi, Kramer, Kuper, Lambropoulos, Law, van Leeuwen, Lemaitre, Loose,
  Maat, Macario, Markoff, Masters, McFadden, McKay-Bukowski, Meijering,
  Meulman, Mevius, Middelberg, Millenaar, Miller-Jones, Mohan, Mol, Morawietz,
  Morganti, Mulcahy, Mulder, Munk, Nieuwenhuis, van Nieuwpoort, Noordam,
  Norden, Noutsos, Offringa, Olofsson, Omar, Orr{\'{u}}, Overeem, Paas,
  Pandey-Pommier, Pandey, Pizzo, Polatidis, Rafferty, Rawlings, Reich,
  de~Reijer, Reitsma, Renting, Riemers, Rol, Romein, Roosjen, Ruiter, Scaife,
  van~der Schaaf, Scheers, Schellart, Schoenmakers, Schoonderbeek, Serylak,
  Shulevski, Sluman, Smirnov, Sobey, Spreeuw, Steinmetz, Sterks, Stiepel,
  Stuurwold, Tagger, Tang, Tasse, Thomas, Thoudam, Toribio, van~der Tol, Usov,
  van Veelen, van~der Veen, ter Veen, Verbiest, Vermeulen, Vermaas, Vocks,
  Vogt, de~Vos, van~der Wal, van Weeren, Weggemans, Weltevrede, White,
  Wijnholds, Wilhelmsson, Wucknitz, Yatawatta, Zarka, Zensus, \& van
  Zwieten}]{vanHaarlem2013}
van Haarlem, M.~P., Wise, M.~W., Gunst, A.~W., {et~al.} 2013, Astronomy {\&}
  Astrophysics, 556, A2

\bibitem[{{van Weeren} {et~al.}(2016{\natexlab{a}}){van Weeren}, {Williams},
  {Hardcastle}, {Shimwell}, {Rafferty}, {Sabater}, {Heald}, {Sridhar},
  {Dijkema}, {Brunetti}, {Br{\"u}ggen}, {Andrade-Santos}, {Ogrean},
  {R{\"o}ttgering}, {Dawson}, {Forman}, {de Gasperin}, {Jones}, {Miley},
  {Rudnick}, {Sarazin}, {Bonafede}, {Best}, {B{\^i}rzan}, {Cassano},
  {Chy{\.z}y}, {Croston}, {Ensslin}, {Ferrari}, {Hoeft}, {Horellou}, {Jarvis},
  {Kraft}, {Mevius}, {Intema}, {Murray}, {Orr{\'u}}, {Pizzo}, {Simionescu},
  {Stroe}, {van der Tol}, \& {White}}]{VanWeeren2016a}
{van Weeren}, R.~J., {Williams}, W.~L., {Hardcastle}, M.~J., {et~al.}
  2016{\natexlab{a}}, \apjs, 223, 2

\bibitem[{{van Weeren} {et~al.}(2016{\natexlab{b}}){van Weeren}, {Brunetti},
  {Br{\"u}ggen}, {Andrade-Santos}, {Ogrean}, {Williams}, {R{\"o}ttgering},
  {Dawson}, {Forman}, {de Gasperin}, {Hardcastle}, {Jones}, {Miley},
  {Rafferty}, {Rudnick}, {Sabater}, {Sarazin}, {Shimwell}, {Bonafede}, {Best},
  {B{\^i}rzan}, {Cassano}, {Chy{\.z}y}, {Croston}, {Dijkema}, {En{\ss}lin},
  {Ferrari}, {Heald}, {Hoeft}, {Horellou}, {Jarvis}, {Kraft}, {Mevius},
  {Intema}, {Murray}, {Orr{\'u}}, {Pizzo}, {Sridhar}, {Simionescu}, {Stroe},
  {van der Tol}, \& {White}}]{VanWeeren2016}
{van Weeren}, R.~J., {Brunetti}, G., {Br{\"u}ggen}, M., {et~al.}
  2016{\natexlab{b}}, \apj, 818, 204

\bibitem[{Venturi {et~al.}(2007)Venturi, Giacintucci, Brunetti, Cassano,
  Bardelli, Dallacasa, \& Setti}]{Venturi2007}
Venturi, T., Giacintucci, S., Brunetti, G., {et~al.} 2007, Astronomy {\&}
  Astrophysics, 463, 937

\bibitem[{Venturi {et~al.}(2008)Venturi, Giacintucci, Dallacasa, Cassano,
  Brunetti, Bardelli, \& Setti}]{Venturi2008}
Venturi, T., Giacintucci, S., Dallacasa, D., {et~al.} 2008, Astronomy and
  Astrophysics, 484, 327

\bibitem[{Verheijen {et~al.}(2008)Verheijen, Oosterloo, van Cappellen, Bakker,
  Ivashina, van~der Hulst, Minchin, \& Momjian}]{Verheijen2008}
Verheijen, M. a.~W., Oosterloo, T.~a., van Cappellen, W.~a., {et~al.} 2008, AIP
  Conference Proceedings, 1035, 265

\bibitem[{Vernstrom {et~al.}(2015)Vernstrom, Norris, Scott, \&
  Wall}]{Vernstrom2015}
Vernstrom, T., Norris, R.~P., Scott, D., \& Wall, J.~V. 2015, Monthly Notices
  of the Royal Astronomical Society, 447, 2243

\bibitem[{Zandanel {et~al.}(2013)Zandanel, Pfrommer, \& Prada}]{Zandanel2013}
Zandanel, F., Pfrommer, C., \& Prada, F. 2013, Monthly Notices of the Royal
  Astronomical Society, 438, 124

\end{thebibliography}

\end{document}